\documentclass{article}
\pdfoutput=1
\usepackage[latin1]{inputenc}
\usepackage[english]{babel}
\usepackage{amsmath}
\usepackage{amssymb,amsfonts}
\usepackage[square,sort,comma,numbers]{natbib}
\usepackage{color}
\usepackage{array}
\usepackage{supertabular}
\usepackage{multirow}
\usepackage{placeins}
\usepackage{tabu}
\usepackage{longtable}
\usepackage{hhline}
\usepackage{microtype} 
\usepackage{geometry}
\usepackage[linkcolor={darkBlue},citecolor={darkBlue},urlcolor={darkBlue}]{hyperref}
\hypersetup{pdftex, colorlinks=true, linkcolor=blue, citecolor=blue, filecolor=blue, urlcolor=black, 
pdftitle={Introduction to the Case Management Model and Notation (CMMN)}, 
pdfauthor= {Mike A. Marin}, 
pdfsubject={Case Management Model and Notation}, 
pdfkeywords={CMMN} {Case Management Model and Notation} {Case Management Modeling}}
\usepackage[pdftex]{graphicx}

\title{Introduction to the Case Management Model and Notation (CMMN)}
\author{Mike A. Marin\\
University of South Africa\\
IBM Analytics Group\\
\texttt{mmarin@acm.org}}
\date{\today}
\begin{document}
\maketitle

\begin{abstract}
This is a short tutorial of the Case Management Model and Notation (CMMN) version 1.0. 
It is targeted to readers with knowledge of basic process or workflow modeling. 
It covers the complete CMMN notation. 
A simple \textit{complaints process} is used to demonstrate the notation.
At the end of the tutorial the reader will be able to understand and create CMMN models.
An appendix summarizing the notation is included for reference purposes.
An interactive but shorter version of this tutorial is available online at \url{http://cmmn.byethost4.com}.
\end{abstract}

\section{Introduction}
Case management is a type of business process technology that does not use control flow to describe the process. The
case (case file or case folder) is the main concept, and it contains all the data and information about the process.
Case management is about empowering workers by providing them with access to all the information concerning the case
and giving them discretion and control on how a case evolves. Case management it is not about the process, it is about
the workers.

In a traditional workflow or process system the designer encodes the business goal to be accomplished in the model.
Thereafter the system is responsible for the business goal and it uses the workers to achieve that goal. In a case
management system, on the other hand, the workers are responsible for the business goal and they use the system as a
tool to accomplish that goal. That it is why case management relies more in the worker's judgment than in control flow.

For this tutorial we will use the Case Management Model and Notation (CMMN) version 1.0 \cite{Omg2014CMMN} to model a fictitious
\textit{complaints process}. The fictitious \textit{complaints process} is a common process in
customer service departments. We will assume this company sells very specialized and expensive products. Due to the
nature of the products the company also provides services (like installation and configuration) for those products. The
company customer service department has a group of highly skilled workers that deal with complaints. The goal of
modeling the \textit{complaints process} is to standardize it to provide guidelines and support to the
customer service workers in charge of that process. In addition, the case management \textit{complaints
process} will allow the company to improve customer service and better track service level agreements (SLA) in the
complaints process.

\subsection{Case Management Model and Notation (CMMN)}

The Case Management Model and Notation (CMMN) version 1.0 was created by the Object Management Group (OMG) and published
in 2014. It is a complementary notation to the Business Process Model and Notation (BPMN) \cite{Omg2013BPMN} which focus on control flow
to describe business processes. We will say that CMMN is declarative in which you describe `what' is allowed and
disallowed in the process; versus BPMN that is imperative in which you describe `how' to do the process. BPMN, CMMN,
and the Decision Model and Notation (DMN) are the three OMG business modeling notations.

The case is the main concept in CMMN, and it is similar to a process. A case contains a case file (i.e. case data
container) and it is described by a case plan (i.e. a model or diagram). We start by introducing the case plan for our \textit{complaints process}.
\section{Case Plan}
In CMMN, a model (i.e. diagram) may have multiple cases, and each case is described by a \texttt{case
plan}. For our \textit{complaints process} we will use a single case, so we will have a single
\texttt{case plan}. The complete description of the case being modeled is described inside its
\texttt{case plan}. However, because of the nature of case management not all the work that happens in
a case is modeled. In particular the interaction of case workers with the case and case data may not be modeled, or be
just partially modeled. In most situations, we don't model how the data gets into the case. Data can be added, removed,
modified by the case workers at any time during the processing of a case without the need of modeling it.

We start our model by adding a \texttt{case plan} and we will name that case
\textit{complaint}.

During modeling, we design one or more \texttt{case plans}. Eventually, the
\texttt{case plan} will be executed and it will be called a case instance. So, a
\texttt{case plan} is similar to a class (i.e. type), and a case instance is an object of that class.

\begin{center}
\includegraphics[width=4.5cm]{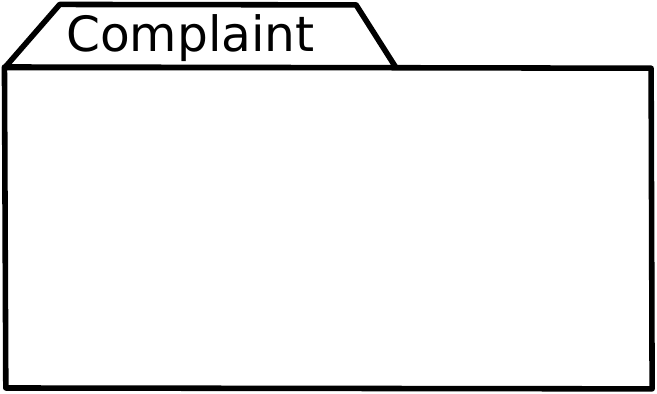}
\end{center}

Although the correct term is \texttt{case plan}, and the word ``case'' is
overloaded (sometimes it refers to a case instance, or the data in the case, or to the process, etc.), for this
tutorial we will use~\texttt{case} to mean \texttt{case plan}.

\subsection{Roles}
In CMMN, roles are defined at the \texttt{case }(\texttt{case plan}) level. However,
CMMN does not provide a graphical notation for roles, and again not everything a role does is modeled. In particular, a
role may be allowed to do case planning (which we will discuss later) that may not be explicitly modeled. A role may
also add, create, modify, or remove data and documents from the case file which may not be explicitly modeled either.

In the \textit{complaints process}, we will have a supervisor, a product specialist, an investigator, etc. In
most situations, we see a case owner role, which in our \textit{complaints process} will be the case worker
who is responsible for managing the complaint. This role decides which tasks must be performed or not in the case
instance for this particular customer.
\section{Case plan items}
An executing model (a case instance) is implementing a plan, and so, its elements are called
\texttt{case plan items}, because they are part of that plan. The \texttt{case plan
items} are \texttt{tasks}, \texttt{stages}, \texttt{milestones}, and
\texttt{event listeners} that we will cover as part of this tutorial.
\subsection{Milestones}

\texttt{Milestones} represent accomplishments during the execution of the case instance. Due to the
large variations between case instances, \texttt{milestones} are important in understanding the
progress of a particular case instance. For our \textit{complaints process}, we will include a few
\texttt{milestones},

\begin{itemize}
\item \textit{Received} to indicate each time we receive customer information. 

\begin{center}
 \includegraphics[width=1.6cm]{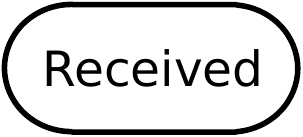} 
\end{center}

\item \textit{Exceed SLA} for those situations in which the case exceed the predefined SLA. 

\begin{center}
 \includegraphics[width=1.6cm]{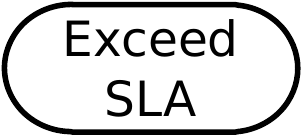}
\end{center}

\item \textit{Completed} to indicate when a product specialist produces a product report. 

\begin{center}
 \includegraphics[width=1.6cm]{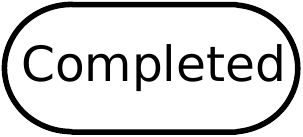} 
\end{center}

\item \textit{Fraud} to indicate when a fraud investigation has started. 

\begin{center}
 \includegraphics[width=1.6cm]{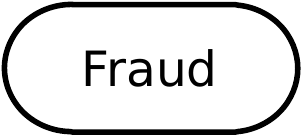} 
\end{center}

\end{itemize}

\subsection{Tasks}

A \texttt{task} represents the execution of actual work. There are four types of
\texttt{tasks}, namely \texttt{non-blocking human task},
\texttt{blocking human task}, \texttt{case task}, and
\texttt{process task}. The type of \texttt{task} is indicated by an icon in the upper
left corner of the \texttt{task} shape. A~\texttt{task} may have a manual activation
decorator (as we will discuss later), which means that a case worker must decide if the task should be executed or not.

In our \textit{complaints process}, we will add a \texttt{non-blocking human task} to hand
over an assignment to a \textit{product specialist.} We will use that \texttt{task} when
executing a \textit{product complaint}, which is one of the types of complaints in our
\textit{complaints process}. We will discuss the \textit{service complaint} later.

It is the responsibility of the \texttt{product specialist} to provide a \textit{report}, so
we don't need to wait until he completes the \texttt{task}. \texttt{Non-blocking
human tasks} are handed out to a case worker (therefore the little hand icon) and as soon as it is claimed by a case worker, it will be considered complete. 

\begin{center}
\includegraphics[width=2cm]{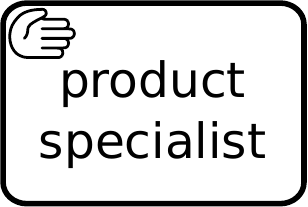} 
\end{center}

We will add a \texttt{case task} to revert a customer payment. This \texttt{task} will create another case to revert the payment. 

\begin{center}
\includegraphics[width=2cm]{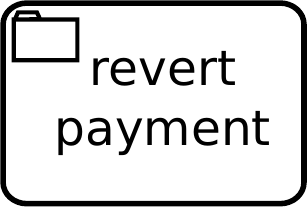} 
\end{center}

We will add a \texttt{task} to inform the customer of the outcome of the case. This will be a
\texttt{blocking human task}. \texttt{Blocking human tasks} are executed by a case
worker and they must be explicitly completed by the worker. 

\begin{center}
\includegraphics[width=2cm]{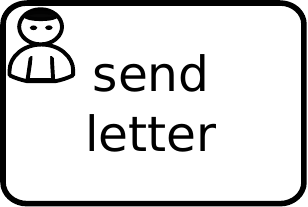} 
\end{center}

Later we will also add a \texttt{process task}. A \texttt{process task} is modeled and
executed using imperative process technology like BPMN. This allows organizations to reuse processes.
\subsection{Stages}

In our fictitious company, there are at least two types of complaints, one for products and one for services. Therefore,
our \textit{complaints process} will need to deal with the two types of complaints, namely
\textit{product complaints} and \textit{service complaints}. We have two options to model the
differences. We could create two different cases, one for each type of complaint, or alternatively we can use
\texttt{stages}. \texttt{Stages} are containers similar to sub-processes in other
workflow or process notations. They are used to manage the complexity of the model by decomposing it into manageable
sets.

For our \textit{complaints process}, we will be sharing most of the data for both types of complaints. In
addition, some situations may involve both service and product complaints. Therefore, \texttt{stages}
make more sense than creating completely separate cases. So, we will add two \texttt{stages} one for
product complaints and another for service complaints.

\begin{center}
 \includegraphics[width=2.2cm]{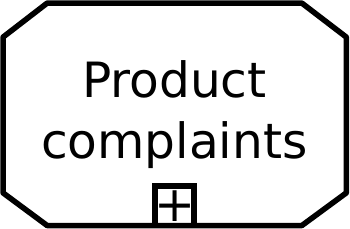} ~~~~~~~~~~  \includegraphics[width=2.2cm]{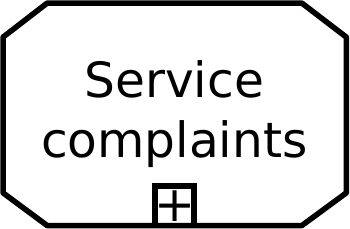} 
\end{center}

\section{Case file}
In CMMN, each case instance contains a single \texttt{case file} (also called a case folder, or just
the case), and case workers have access to all the data in that \texttt{case file}. Case workers can
add, remove, and modify data in the \texttt{case file} even if they are not executing any
\texttt{task} in the case, as long as they have sufficient privileges. The data in the
\texttt{case file} is called \texttt{case }\texttt{file items}.
\subsection{Case file items}

All data and data structures are called \texttt{case file items}. All the \texttt{case
file items} are stored in the \texttt{case file}. \texttt{Case file items} are used
to represent all kinds of data, including a data value in a database, a row in a database, a document, a spreadsheet, a
picture, a video, a voice recording, etc. In addition to basic data, \texttt{case file items} can also
represent containers, including, a directory, a folder, a set, a stack, a list, etc.

As with most case management applications, we will not model all the data required for the \textit{complaints
process}, but we will model the following data,

\begin{itemize}
\item The \textit{report}. As mentioned before, in some situations the \textit{product specialist}
will produce a \textit{report} that will satisfy the \textit{completed}
\texttt{milestone}. In out fictitious company, the \textit{report} is a document that
follows a particular template. 

\begin{center}
 \includegraphics[height=1.3cm]{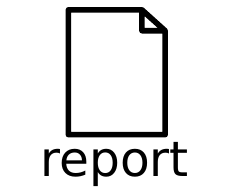} 
\end{center}

\item The \textit{input} folder may contain emails, documents, pictures, recording of customer calls, etc.
All the data submitted by the customer will be collected in a folder called \textit{input}.

\begin{center}
 \includegraphics[height=1.3cm]{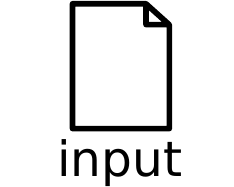} 
\end{center}

\item The \textit{resolution} document is produced as part of the case and indicates the outcome of the
complaints investigation. 

\begin{center}
 \includegraphics[height=1.3cm]{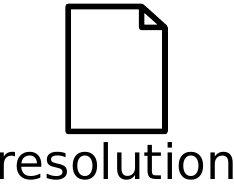} 
\end{center}

\item Finally, a customer can cancel the complaint by calling or sending a notification. That will be represented by the
\textit{cancel} \texttt{case file item}. 

\begin{center}
 \includegraphics[height=1.3cm]{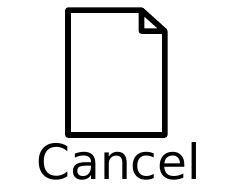} 
\end{center}

\end{itemize}


So far, we have created a set of artifacts in the model (i.e. \texttt{case plan},
\texttt{milestones}, \texttt{tasks}, \texttt{stages}, and
\texttt{case file items}), but we have not modeled how they are related. CMMN is declarative, and it
reacts to changes on the environment and case worker decisions. \texttt{Entry criteria} and
\texttt{exit criteria} will allow us to model changes in the case environment.
\section{Criteria}
Criterion allow us to describe when a \texttt{task}, \texttt{stage}, or
\texttt{milestone} should be available for execution (\texttt{entry criteria}), or
when a \texttt{case} (\texttt{case plan}), \texttt{stage}, or
\texttt{task} should terminate abnormally (\texttt{exit criteria}). Criteria has the
following two optional parts,

\begin{itemize}
\item One or more trigger events (called \texttt{onParts}). These are events that will satisfy the
evaluation of the \texttt{entry criteria} or \texttt{exit criteria}. Events that
emanate from other CMMN elements can be visualized by an optional \texttt{connector} (a dotted line).
However, the visualization is optional and does not describe the type of event. 
\item A Boolean expression (called \texttt{ifPart}). This expression must evaluate to true for the
\texttt{entry criteria} or \texttt{exit criteria} to be satisfied. 
\end{itemize}
We can think of the criteria forming a sentence as follows,

\begin{center} 
\small
\texttt{(}[ \texttt{on} {\textless}Event 1{\textgreater}[,
\texttt{on} {\textless}Event 2{\textgreater}[, . . .]] ]\texttt{) AND (}[
\texttt{if} {\textless}Boolean condition{\textgreater} ]\texttt{)}
\end{center} 

Where square brackets ([ ]) indicate optional parts of the sentence, and angled brackets ({\textless} {\textgreater})
are place holders to be replaced. Looking at the sentence, we can see why the event is called the
\texttt{onPart}, and the Boolean condition is called the \texttt{ifPart}. Note that
both the \texttt{onPart} and the \texttt{ifPart} are optional in the sentence, but
for it to make sense at least one of them must be present.
\subsection{Entry criteria}

An \texttt{entry criterion} ${\lozenge}$  describes the condition that must be satisfied for the
\texttt{stage}, \texttt{task}, or \texttt{milestone} to be available
for execution. \texttt{Stage}, \texttt{task}, or \texttt{milestones}
without \texttt{entry criteria} will be available for execution as soon as they are created.

In the \textit{complaints process}, both \texttt{stages} \textit{product
complaints} and \textit{service complaints} need an \texttt{entry criteria}, because they
can only execute if the complaint is of their type. In most cases, only one of the two \texttt{stages}
will execute, although in some situations the complaints may involve both \texttt{stages}.

\begin{center} 
 \includegraphics[width=2.2cm]{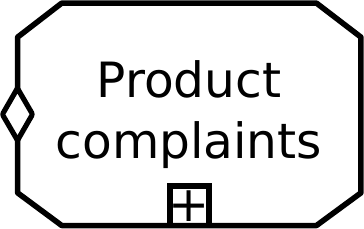} ~~~~~~  \includegraphics[width=2.2cm]{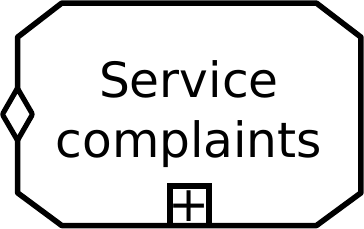} 
\end{center} 

The \texttt{entry criteria} can be placed anywhere in the border of the
\texttt{stage}, \texttt{task}, or \texttt{milestone}. For our
example, we said that a \textit{product specialist} may create a \textit{report} and that
creation generates an event that satisfies the \textit{completed} \texttt{milestone}. We
visualize that as follows,

\begin{center}  
\includegraphics[width=4.5cm]{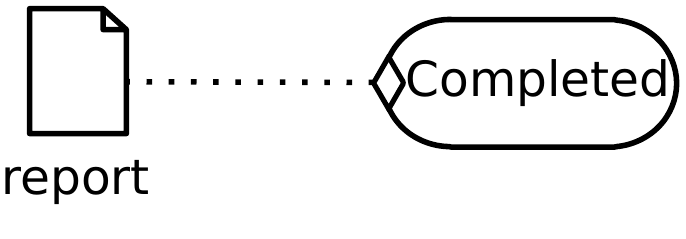} 
\end{center} 
Note that without looking inside the \texttt{entry criteria} of the \textit{completed}
\texttt{milestone}, we cannot tell if it is satisfied by the \textit{report} being created,
updated, or even deleted. In addition, we don't know if there is a condition (\texttt{ifPart}). We
only know there is an event (\texttt{onPart}) from \textit{report} that is used to trigger
the \textit{completed} \texttt{milestone}.


We know the \textit{report} \texttt{case file item} and the \textit{complete}
\texttt{milestone} are part of processing a \textit{product complaint}. We also know the
\textit{product specialist} \texttt{non-blocking task} is also part of processing a
\textit{product complaint}. So, we can expand the \textit{product complaints}
\texttt{stage} to show that, as follows,

Notice that \textit{product specialist} \texttt{task} does not have~
\texttt{entry criteria}, and so, as soon as \textit{product complaints} start executing the
\textit{product specialist} \texttt{task} also start executing. Eventually, the case worker
may create a \textit{report} and that creation event satisfies the \textit{completed}
\texttt{milestone}. 

\begin{center}
\includegraphics[width=5cm]{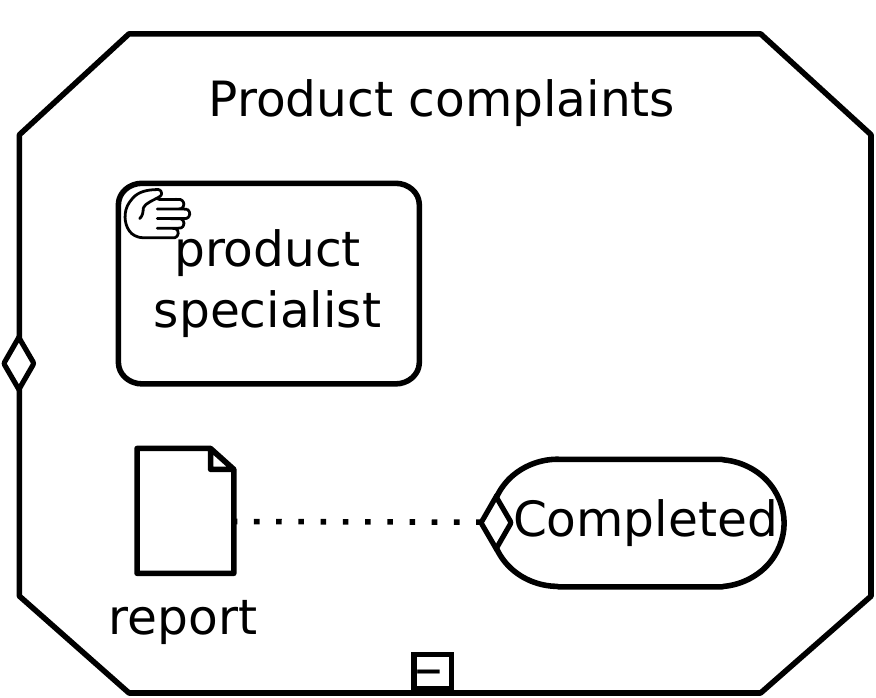} 
\end{center}

In the \textit{complaints process}, the \textit{send letter} \texttt{task}
depends on two events, one from the \textit{received} \texttt{milestone}, and one from the
\textit{resolution} \texttt{case file item}. We require \textit{input} from the
customer, and we have to complete the \textit{resolution} of the case to send the letter to the customer.
That means the \texttt{entry criteria} is waiting for two events (two \texttt{onParts}), and they forms an AND condition. Both events must happen for the \texttt{entry criteria} to be satisfied. 

\begin{center}
\includegraphics[width=5cm]{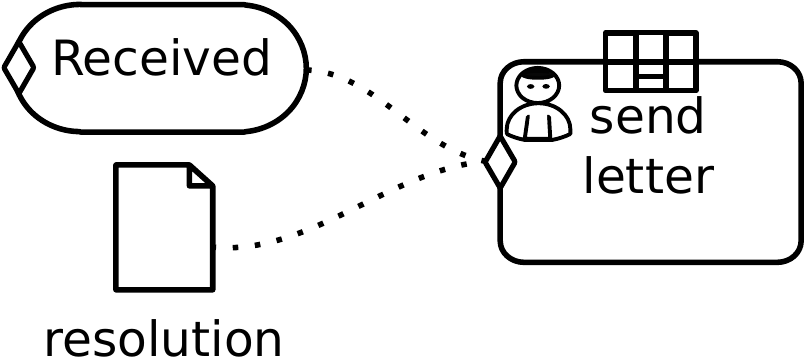} 
\end{center}

Later we will see an example of an OR conditions (hint, use multiple entry conditions).
\subsection{Exit criteria}

An \texttt{exit criterion} ${\blacklozenge}$  is similar to an \texttt{entry
criterion}, but it is used to stop working on the \texttt{stage}, \texttt{task}, or
\texttt{case} (\texttt{case plan}) when it is satisfied.

In the \textit{complaints process}, we will add an \texttt{exit criterion} for the case. In
the situation the customer calls and cancels the complaint, so we need to stop working on the case. We model this
scenario by having a \textit{cancel} \texttt{case file item}, which could be a voice
recording of a customer call, a letter from the customer, or just a flag in the case data.

\begin{center}
\includegraphics[width=4cm]{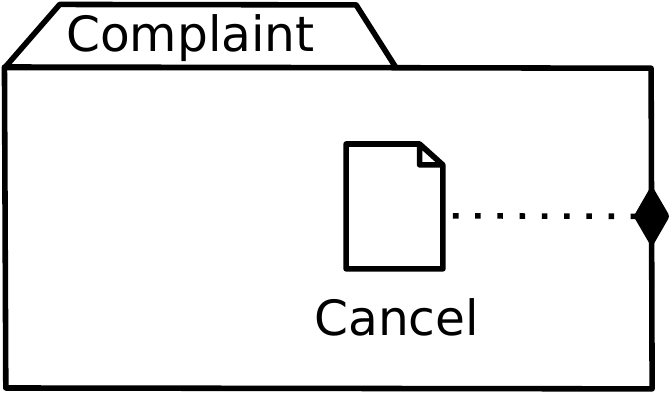}
\end{center}

\subsection[Connecting two criteria]{Connecting two criteria}
It is valid for an \texttt{onPart} to reference another criterion, instead of an event. When the other
criterion is satisfied the \texttt{onPart} will also be satisfied. This is visualized by a connector
($\ldots\ldots\ldots$) between the two criteria. Connecting an \texttt{exit criterion} to an
\texttt{entry criterion} in this way is the only flow control available in CMMN.
\section{Event Listeners}

\texttt{Events listeners} are similar to events in other workflow or BPM notations. For the
\textit{complaints process}, we have a service level agreement (SLA) that we will model using a
\texttt{timer event listener} and a \texttt{milestone}, as follows.

\begin{center}
 \includegraphics[width=4cm]{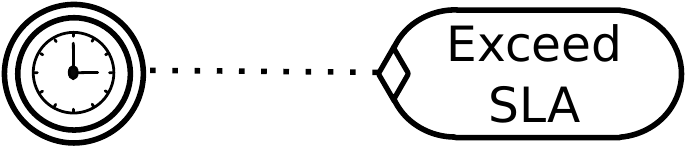} 
\end{center}

We will have a \texttt{human event listener} to provide the supervisor with a way to trigger the
\textit{revert payment} \texttt{task}. There are multiple ways that we can use to give the
supervisor the ability to trigger the \textit{revert payment} \texttt{task}, but for
illustration purposes we will use a \texttt{human event listener}.

\begin{center}
 \includegraphics[width=4cm]{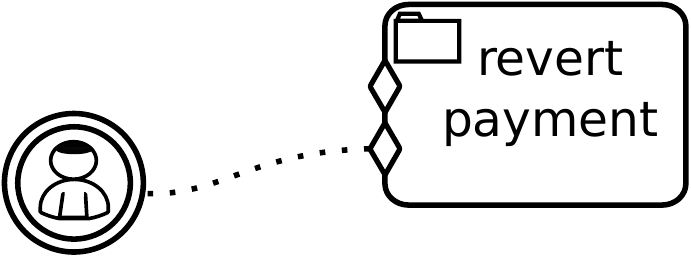} 
\end{center}

Note that \textit{revert payment} \texttt{task} has two \texttt{entry
criteria}, one is triggered by the \texttt{human event listener}, and the other is triggered by
another \texttt{entry criteria}. When a \texttt{task},
\texttt{stage}, or \texttt{milestone} has more than one
\texttt{entry criteria} they form an OR condition. Meaning any of the \texttt{entry
criteria} that is satisfied will start the \texttt{task}, \texttt{stage}, or trigger
the \texttt{milestone}.
\subsection{Plan items standard events}

As we described before, the \texttt{entry criteria} and \texttt{exit criteria} have an
event (\texttt{onPart}) and a condition (\texttt{ifPart}). An executing model (a case
instance) is implementing a plan, and so, \texttt{tasks}, \texttt{stages},
\texttt{milestones} and \texttt{event listeners} are called
\texttt{case plan items}, because they are part of that plan. Every \texttt{case plan
item} generates events that can be used in the \texttt{onPart} of an \texttt{entry
criteria} or \texttt{exit criteria}.

The following table lists the standard events for the different \texttt{case plan items}. There is no
need to memorize this table of events, because modeling tools should have the list available when defining an
\texttt{entry criteria} or \texttt{exit criteria}. The table contains a column
indicating the events that result from a case worker action. As you can see a case worker (in a role with enough
privileges) has a lot of discretion to control and modify the behavior of the \texttt{case plan
items}.

\begin{longtabu}{|l|l|l|l|p{5.5cm}|} \hline
\textbf{Cases} & \textbf{Tasks \& Stages } & \textbf{Events \& }  & \textbf{Case}& \textbf{Description} \\
& & \textbf{Milestones } & \textbf{worker} & \\
{\includegraphics[width=1cm]{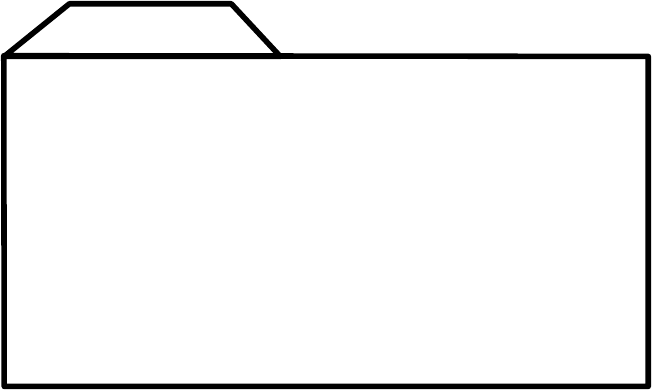}} 
& \includegraphics[width=1cm]{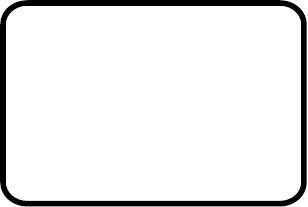} \includegraphics[width=1cm]{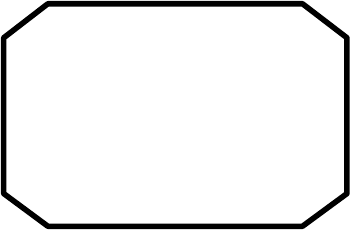}
& \includegraphics[width=0.5cm]{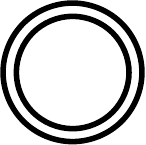}  \includegraphics[width=1cm]{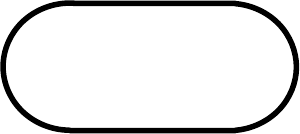}
& & \\
& \includegraphics[width=1cm]{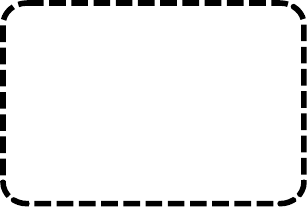} \includegraphics[width=1cm]{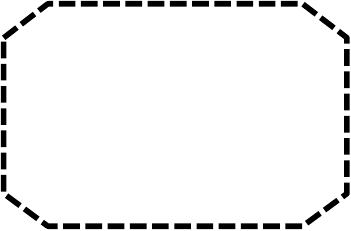} & & &\\\hline
\textbf{create}     & \textbf{create}        & \textbf{create}    &     & element was created\\
           & \textbf{start}         &           &     & started executing\\
           & \textbf{enable}        &           & Yes & enabled after disabled\\
           & \textbf{manualStart}   &           & Yes & manual execution (with manual activation annotator)\\
           & \textbf{disable}       &           & Yes & disable element that has not started execution\\
           & \textbf{reenable}      &           & Yes & reenable (after it was disabled)\\
\textbf{suspend}    & \textbf{suspend}       & \textbf{suspend}   & Yes & suspend element that is executing\\
           & \textbf{resume}        & \textbf{resume}    & Yes & resume execution (after it was suspended)\\
           &               & \textbf{occur}     &     & milestone or event occurs\\
           & \textbf{parentSuspend} &           &     & parent suspended\\
           & \textbf{parentResume}  &           &     & parent resumed \\
\textbf{reactivate} & \textbf{reactivate}    &           & Yes & reactivate after a failure occurred\\
\textbf{complete}   & \textbf{complete}      &           &     & normal completion\\
\textbf{terminate}  & \textbf{terminate}     & \textbf{terminate} & Yes & Manual termination or exit criteria\\
\textbf{fault}      & \textbf{fault}         &           &     & element is in a fault condition\\
\textbf{close}      &               &           & Yes & case is closed\\\hline
\end{longtabu}

\subsection{Case file item standard events}

\texttt{Case file items} (data) also generate events that can be used in an
\texttt{entry criteria} or \texttt{exit criteria}. The following table lists the
standard events for the case file items. Again, there is no need to memorize this table of events, because modeling
tools should have the list available when defining an entry or exit criteria. The table contains a column indicating
the events that result from a case worker action. As you can see a case worker (in a role with enough privileges) has
the ability to add, delete, and modify the data in the case.

\begin{longtabu}{|l|l|l|}\hline
\textbf{Case file item} & \textbf{Case} & \textbf{Description}\\
& \textbf{worker} & \\
\includegraphics[width=0.6cm]{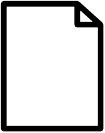} & & \\\hline

\multicolumn{3}{|m{13cm}|}{All type of data (for example, a row in a database, a document, a picture, a video,
etc.) generate the following events}\\\hline
\textbf{create} &
\centering Yes &
Item was created\\\hline
\textbf{replace} &
\centering Yes &
Content of the item has been replaced\\\hline
\textbf{update} &
\centering Yes &
Item has been updated\\\hline
\textbf{delete} &
\centering Yes &
Item has been deleted\\\hline
\textbf{addReference} &
\centering Yes &
A new reference to the item has been added\\\hline
\textbf{removeReference} &
\centering Yes &
a reference to the item has been removed\\\hline
\multicolumn{3}{|m{13cm}|}{Containers (for example, a directory, a folder, a set, a stack, a list, a database
table, etc.) Generate (in addition to the previous events) the following extra events}\\\hline
\textbf{addChild} &
\centering Yes &
a new child has been added to the container\\\hline
\textbf{removeChild} &
\centering Yes &
a child has been removed from the container\\\hline
\end{longtabu}

\section{Planning}

We have defined most of the \textit{complaints process} model. In particular, we have defined most of what
will be in the \texttt{case} (\texttt{case plan}) when a case instance is created.
Putting everything together, this is what we have modeled so far,

\begin{center}
 \includegraphics[width=13cm]{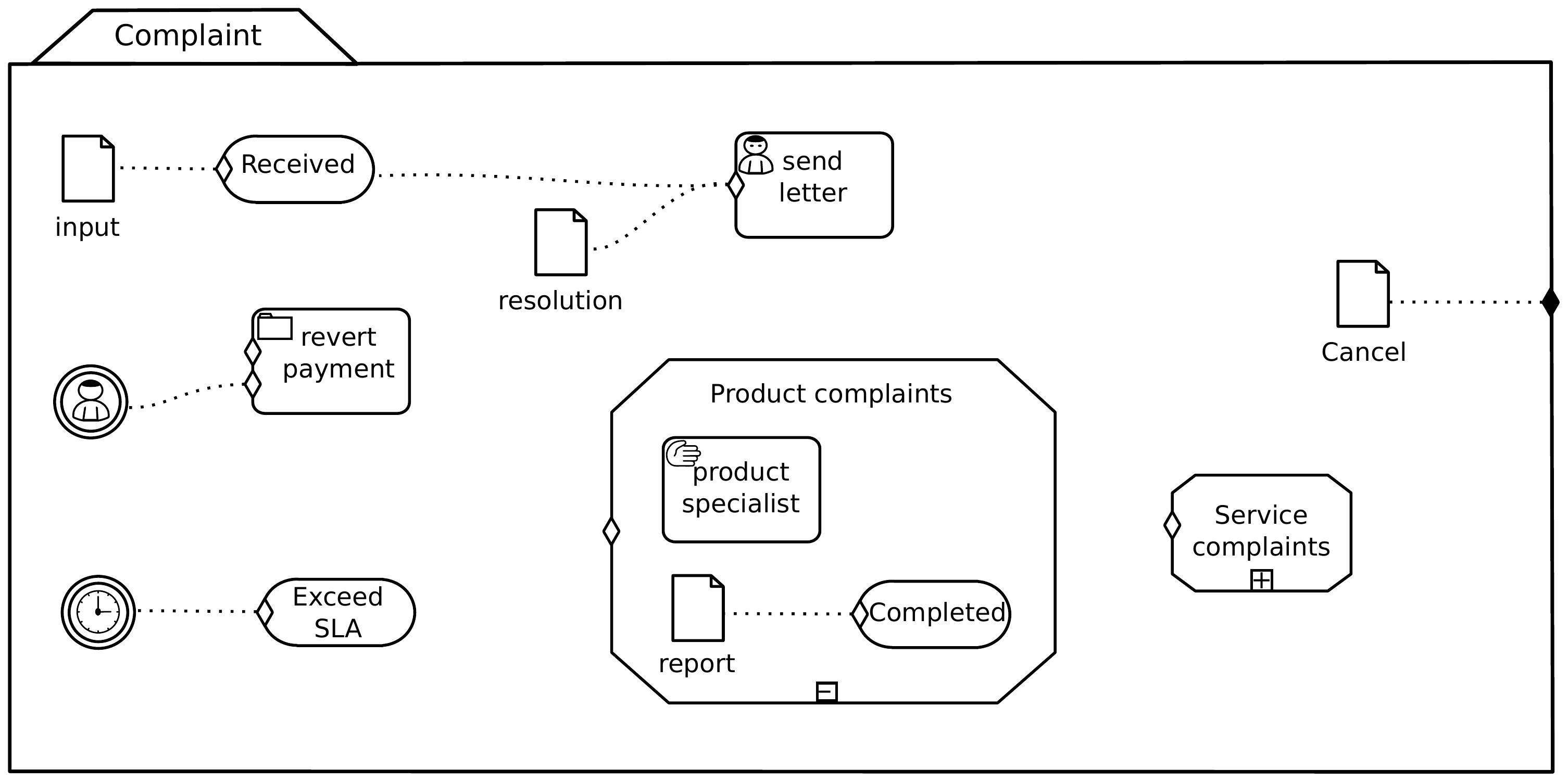} 
\end{center}

This model shows everything that will be in the execution plan, when a case instance is created. But, CMMN provides a
way for case workers to add more elements to the plan. Remember that case workers can disable
\texttt{case plan items}, but they can also add new items to the plan. CMMN has the concept of
\texttt{discretionary items}, which are modeled but are not included in the execution plan of a case
instance. The only way a \texttt{discretionary item} will be added to the plan of a case instance is
when a case worker adds it to the plan. Adding \texttt{discretionary items} to the
\texttt{case} is called planning. \texttt{Discretionary items} have the same shape as
planned items, but using a dashed line, instead of the continuous line used by \texttt{case plan
items}.

In some uncommon situations the case workers in our fictitious company may suspect they may be dealing with a fraudulent
complaint. But, because this situation is not common there is no need to include it in the plan of every case instance.
However, we want to give case workers the ability to add a \textit{Fraud investigation}
\texttt{stage} to the \texttt{case}, so we will add a
\texttt{discretionary stage} as follows,

\begin{center}
\includegraphics[width=5cm]{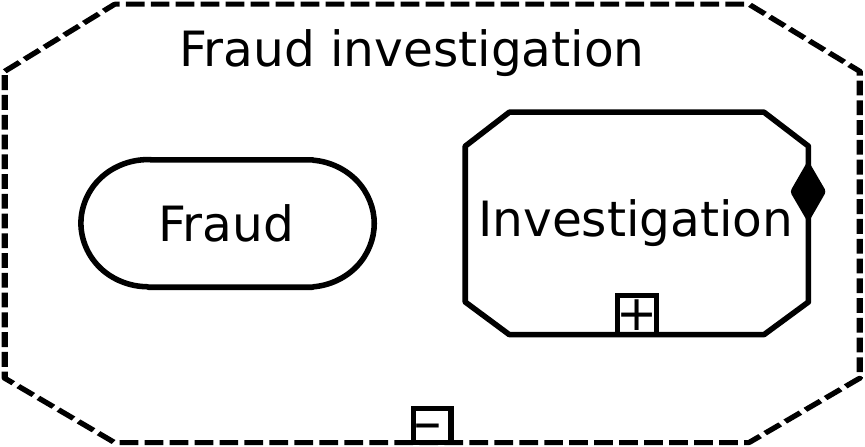} 
\end{center}

Note that \textit{Fraud investigation} is a \texttt{discretionary stage}, and so it is drawn
with a dashed line. Also notice that it contains two non-discretionary items (\textit{Fraud}
\texttt{milestone} and \textit{Investigation} \texttt{stage}). Both, the
\textit{Fraud} \texttt{milestone} and the \textit{Investigation}
\texttt{stage} will start executing as soon as \textit{Fraud investigation} executes,
because they are non discretionary and they don't have any entry criteria. 

\subsection{Planned versus discretionary}
The CMMN specification describes the distinction between the execution plan and planning, with the following diagram,

\begin{center}
 \includegraphics[width=9.5cm]{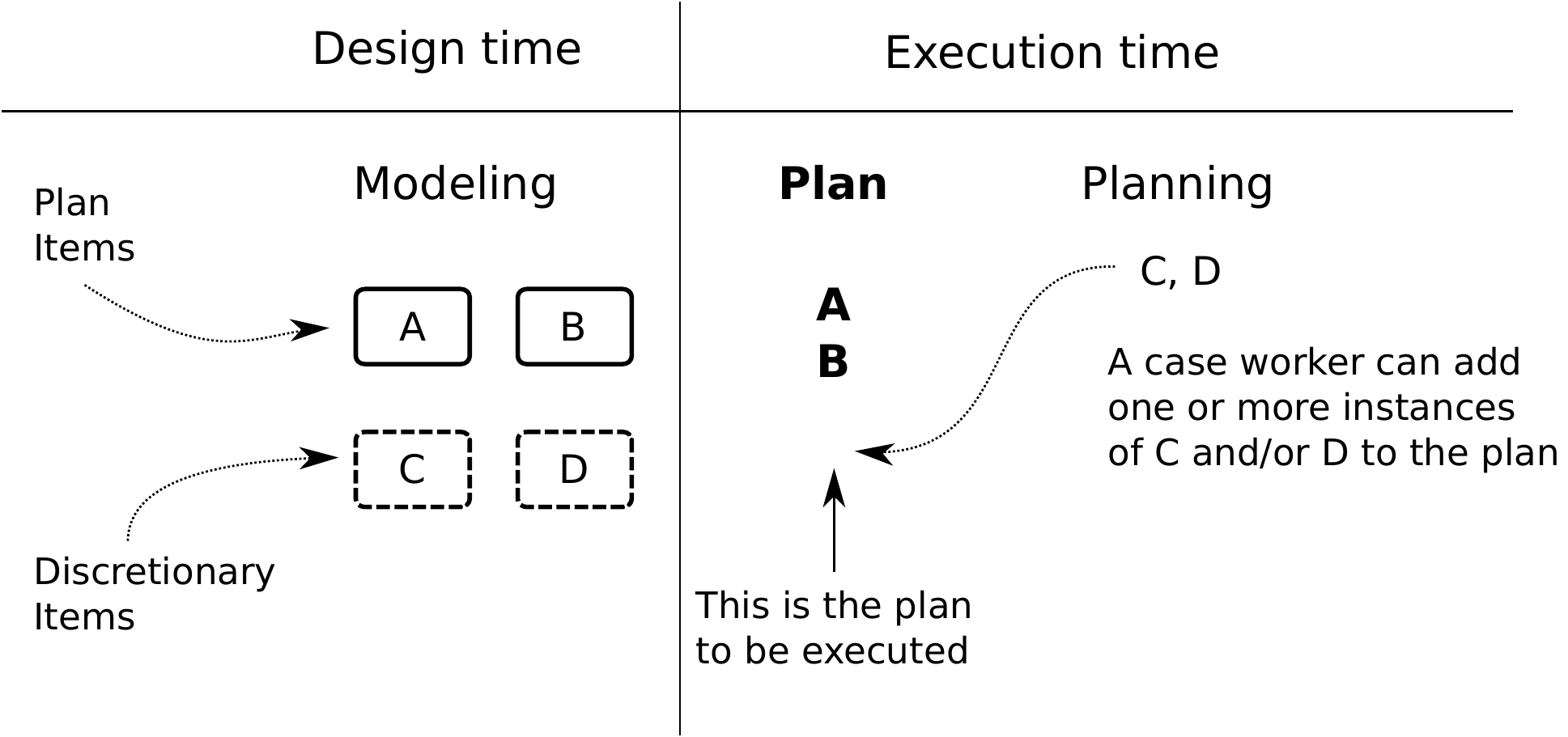} 
\end{center}

At design time, the user models both \texttt{plan items} and \texttt{discretionary
items}. At execution time, a CMMN compliant engine only has in the execution plan the \texttt{plan
items}. However, case workers can add to the plan the \texttt{discretionary items} they consider
necessary for the particular case instance. 

\subsection{Planning table}

\texttt{Planning tables} are used to indicate that planning is allowed in a
\texttt{case} (\texttt{case plan}), \texttt{stage}, or
\texttt{human task}. For planning to be allowed, there must be \texttt{discretionary
stages}, \texttt{discretionary tasks}, or \texttt{plan fragments} in the scope of the
\texttt{case}, \texttt{stage}, or \texttt{human task}. To indicate
that a case worker can do planning a \texttt{planning table} is used. The
\texttt{planning table} can be expanded or collapsed.

\begin{center} 
 \includegraphics[width=0.8cm]{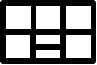} ~~ Expanded
\end{center} 

\begin{center} 
 \includegraphics[width=0.8cm]{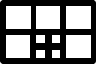} ~~ Collapsed
\end{center} 

In our \textit{complaints process}, there are situations in which the case worker sending the letter to the
customer may feel it is appropriate to provide the customer with a discount. We don't want to encourage the discount
(and so we don't put it in the plan), but we want to give the case worker the ability to add a
\textit{process discount\texttt{ }\texttt{discretionary task}} to the
plan. The icon inside the \textit{process discount\texttt{
}\texttt{discretionary task}} indicates this is a \textit{\texttt{process
task}}, which we have not seen before. So, we add a \textit{process discount}
\texttt{discretionary task}, and we place that \texttt{discretionary task} in the
scope of the \textit{send letter} \texttt{task} by adding a
\texttt{planning table}, as follows

\begin{center}  
\includegraphics[width=5cm]{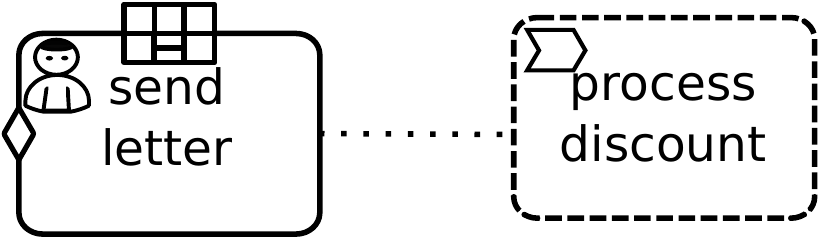} 
\end{center} 
Notice that we draw an expanded \texttt{planning table} in the border of the \textit{send
letter} \texttt{task}. In addition, because the \texttt{planning table} is expanded,
we show the \textit{process discount} \texttt{discretionary task} connected to the send
letter task by a \texttt{connector}. Note the \texttt{connector} in this situation is
different from a \texttt{connector} in an \texttt{entry criteria} or
\texttt{exit criteria}. In this situation, the dotted line is indicating that a case worker executing
the \textit{send letter} \texttt{task} can add a \textit{process discount}
\texttt{discretionary task} to the plan, and the \textit{process discount}
\texttt{discretionary task} will start executing as soon as it is added to the case instance plan.

We will also add a \textit{check safety} \texttt{discretionary task} to our
\textit{product complaints} \texttt{stage}. This \texttt{discretionary
task} will be used in rare situations when a case worker suspects a product safety issue. Therefore, we need to add a
\texttt{planning table} to this \texttt{stage}. We also added other
\texttt{discretionary items} for the \textit{product specialist} to use, but because of
space considerations, we will not show them. These are indicated by the collapsed \texttt{planning
table} in the \textit{product specialist} \texttt{task}. So, our \textit{product
complaints} \texttt{stage} now looks as follows, 

\begin{center}
\includegraphics[width=6cm]{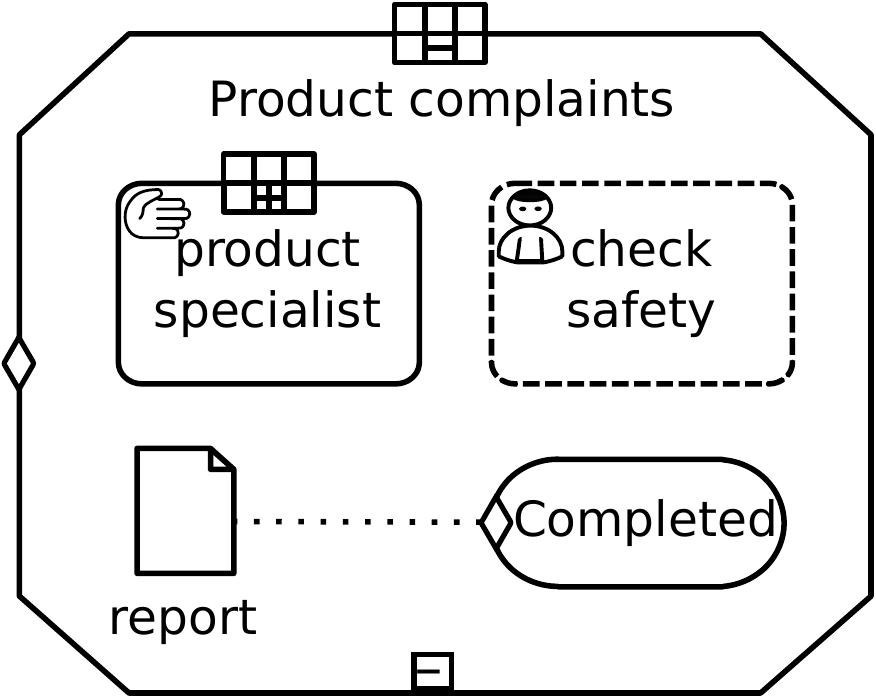} 
\end{center}

\subsection{Plan fragment}
There are situations in which you want to give the case workers the ability to add a set of
\texttt{discretionary items} as a single planning action. \texttt{Plan fragments}
provide a way to do that. A \texttt{plan fragment} is just a grouping mechanism for
\texttt{discretionary items}. In our example, \textit{Audit} is a
\texttt{plan fragment} that can be started by a manager with the ability to add
\texttt{discretionary items} to the case plan.

\begin{center}
\includegraphics[width=2cm]{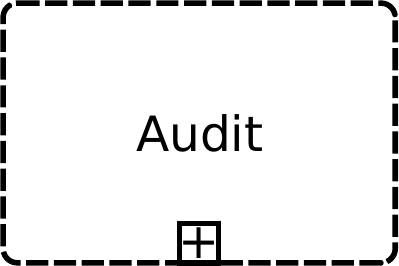} 
\end{center}

Note that \textit{Audit} is collapsed $\boxplus$, and so, we
don't see the \texttt{discretionary items} it contains. Both, \textit{Audit} and
\textit{Fraud investigation} will be at the case level scope, so we need to add a
\texttt{planning table} to the \textit{complaints} \texttt{case} also.

\section{Decorators}

\texttt{Case plan items} and \texttt{discretionary items} can be annotated to indicate
certain characteristics of the item. There are four decorators,

\subsection{Auto complete decorator}
The \texttt{Auto complete} decorator ${\blacksquare}$ 
indicates that the \texttt{stage} or \texttt{case} (\texttt{case
plan}) will complete when all the required \texttt{case plan items} are completed. If the decorator is
not present, the \texttt{stage} or \texttt{case} requires manual completion by a case
worker after all the required \texttt{case plan items} have completed. You will use this decorator
when there are no \texttt{discretionary items} and you want the system to complete the
\texttt{case} or \texttt{stage} as soon as all the \texttt{case plan
items} have completed. However, in most situations you will want a case worker deciding if a
\texttt{stage} or \texttt{case} should be completed. Because there may be
\texttt{discretionary items} or other reasons to avoid prematurely closing the
\texttt{case} or \texttt{stage}.

We will place an \texttt{auto complete} decorator in the \textit{Fraud investigation}
\texttt{discretionary stage}, because we want it to complete as soon as both the
\textit{Fraud} \texttt{milestone} and the \textit{Investigation}
\texttt{stage} complete.

\begin{center}
 \includegraphics[width=6cm]{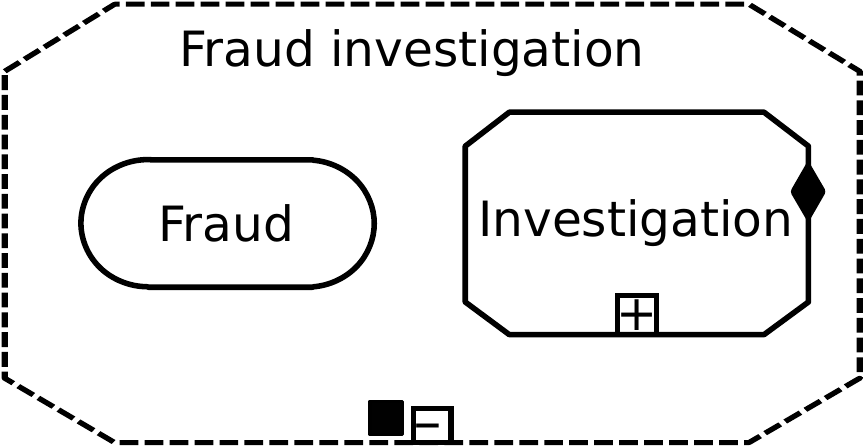} 
\end{center}

\subsection{Manual activation decorator}
The \texttt{Manual activation} decorator $\vartriangleright$ indicates that the \texttt{stage} or \texttt{task} must be manually initiated
after the \texttt{entry criteria} has been satisfied. If the decorator is not present, the
\texttt{stage} or \texttt{task} will automatically start executing when one of the
\texttt{entry criteria} is satisfied. It is important to provide case workers veto power over the case
management system. There are situations, in which an \texttt{entry criterion} is satisfied, but you
want to give case workers the ability to decide if the \texttt{task} or
\texttt{stage} really needs to be executed and when to start the execution. In our
\textit{complaints process}, we have a \textit{revert payment} \texttt{task}
that will be ready for execution if one of its two entry criterion is satisfy. However, we want to be sure that a case
worker verify if \textit{revert payment} should be executed. 

\begin{center}
 \includegraphics[width=5cm]{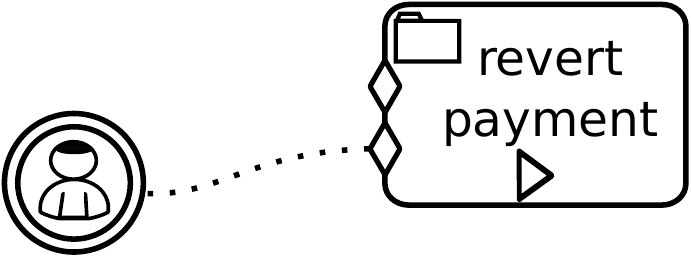} 
\end{center}


\subsection{Required decorator}
\texttt{Required decorator} \texttt{!} indicates that a
\texttt{stage}, \texttt{task}, or \texttt{milestone} must be
executed for the scope (\texttt{stage} or \texttt{case}) to complete. Note that every
\texttt{case plan item} that starts executing must also complete for the enclosing
\texttt{case} or \texttt{stage} to complete. Therefore, as soon as a
\texttt{case plan item} starts execution it becomes required. \texttt{Discretionary
items}, by definition are not part of the plan, so they cannot have a \texttt{required} decorator.
However, when a \texttt{discretionary item} is added to the plan by a case worker and starts execution
it must complete execution for the enclosing \texttt{case} or \texttt{stage} to
complete. In our \textit{complaints process}, we will mark the \textit{send letter}
\texttt{task} as \texttt{required}, because we want to be sure we notify the customer
of the outcome of the complaints. 

\begin{center}
 \includegraphics[width=5cm]{tu-human-expanded-table.pdf} 
\end{center}

\subsection{Repetition decorator}
\texttt{Repetition decorator} \texttt{\#}  indicates the
\texttt{stage}, \texttt{task}, or \texttt{milestone} can be repeated
multiple times. Only \texttt{stages}, \texttt{tasks}, or
\texttt{milestones} with at least one \texttt{entry criteria} can have the
\texttt{repetition} decorator. In the \textit{complaints process} we will mark the
\textit{Received} \texttt{milestone} with the \texttt{repetition}
decorator, because the customer may send multiple different documents or pieces of information that will be stored in
the \textit{input} \texttt{case file item}, and we want the \textit{Received}
\texttt{milestone} to occur each time we receive something from the customer. 

\begin{center}
 \includegraphics[width=4.5cm]{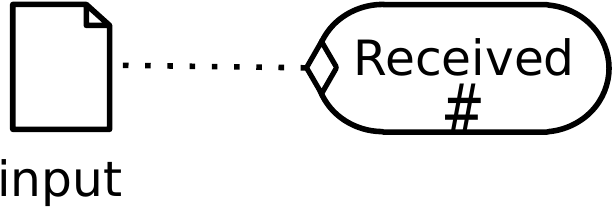} 
\end{center}

\section{Connectors}
We have already seen the two situations in which \texttt{connectors} (. . . . . .) are used. First,
they are used to visualize the event (\texttt{onPart}) of \texttt{entry criteria} or
\texttt{exit criteria}. In this situation the \texttt{connector} is optional. Second,
\texttt{connectors} are used to visualize the \texttt{discretionary items} associated
with a \texttt{human task}{}'s \texttt{planning table}, in which case the
\texttt{connector} connects a \texttt{human task} with a
\texttt{planning table} to the \texttt{discretionary items} in the
\texttt{planning table}.
\section{Complete model}

The complete \textit{complaints} case model, including \texttt{discretionary items} and
decorators, looks as follows,

\begin{center}
 \includegraphics[width=12.5cm]{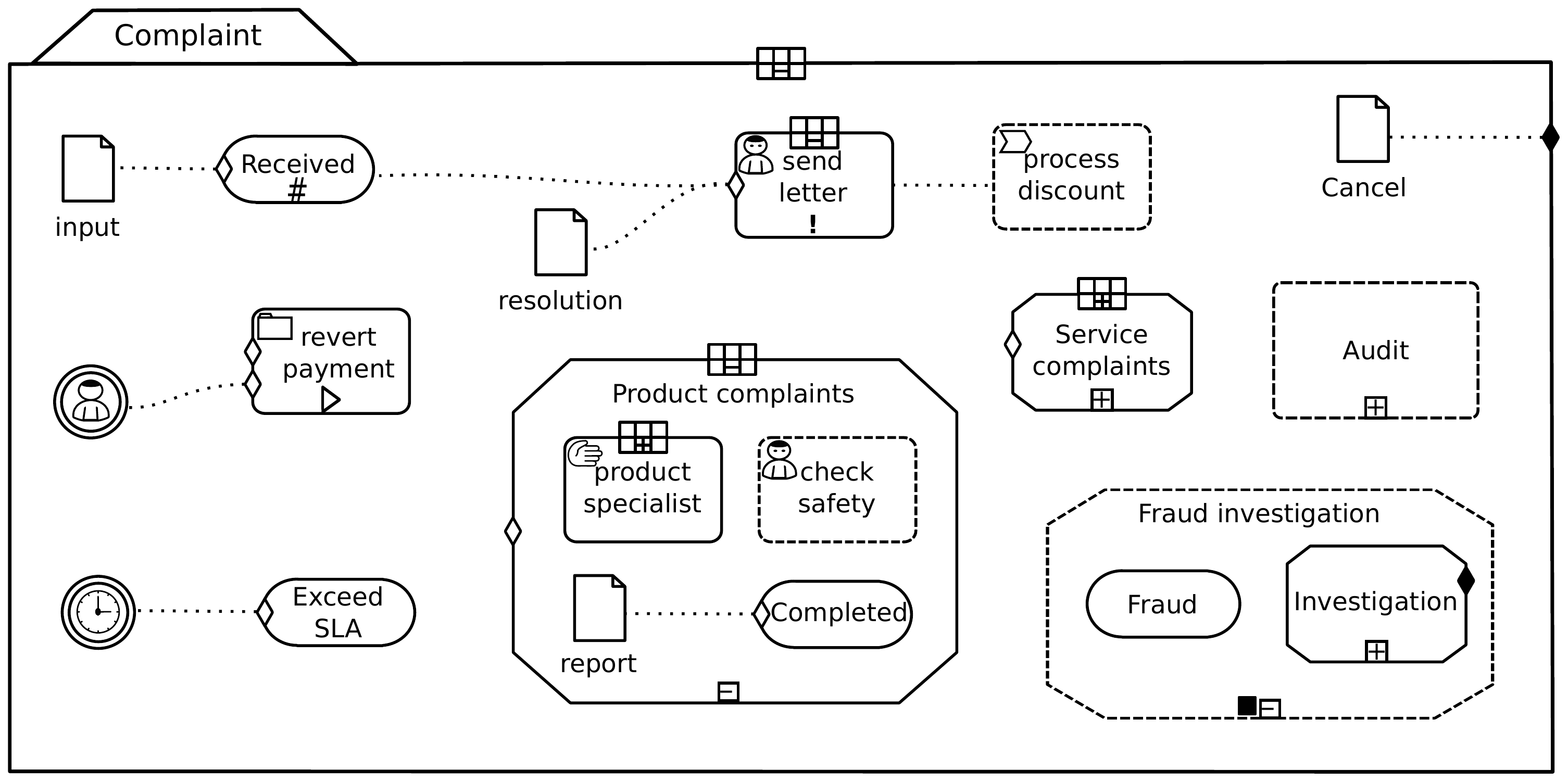} 
\end{center}

There are a few things we left out of the model, namely \textit{Service complaints} and
\textit{Audit} are collapsed, so we did not show their content; \textit{Product complaints}
\texttt{stage} and \textit{product specialist} \texttt{task} contain
collapsed \texttt{planning tables}, so we don't show the \texttt{discretionary items}
they contain.

Note that the full CMMN model for the complaint process may look disconcerting to a person used to workflow or process
models, because not everything is modeled, and because most of what is modeled can be disabled at execution time by
case workers.
\subsection{Case Worker Actions}

As we have described before, CMMN gives case workers a lot of control over the execution of a case instance. Although
case worker privileges are controlled using roles, it is important to distinguish two types of case workers.

\begin{itemize}
\item Case workers executing \texttt{tasks} in the case. These workers may have limited privileges and
will have similar characteristics to workers in other process or workflow technologies. 
\item Case workers controlling the case. These are sometimes referred as knowledge workers. In the
\textit{complaints process} we have a case worker dealing with the customer and in charge of the case. Some
of the activities that case workers controlling the case are able to do include, 
\end{itemize}
\texttt{Case planning}. Adding \texttt{discretionary items} to the case plan.

\texttt{Manual activation}. Deciding when a \texttt{task} or
\texttt{stage} should be executed (by manually starting it), or deciding that it should not be
executed for a case instance (by disabling it).

\texttt{Auto complete}. Deciding when a \texttt{stage} or
\texttt{case} without \texttt{auto complete} should be manually completed.

\texttt{Suspend and resume}. Deciding when to suspend or resume execution of the
\texttt{case}, \texttt{task}, \texttt{stage},
\texttt{event listener}, or \texttt{milestone}.

\texttt{Ignoring fault conditions}. Deciding to continue a \texttt{case}, a
\texttt{task}, or a \texttt{stage} that has an error condition.

\texttt{Adding or modifying data on the case}. Adding, creating, replacing, deleting, and modifying
data (\texttt{case file items}) in the case.

\texttt{Closing the case}. Deciding when to close a case, so that nothing else can be done in that
case.
\section{Summary}

Case management looks at a process from the perspective of the case workers, with the goal of enabling them to
efficiently collaborate to achieve a business goal. CMMN achieves that by allowing execution time planning of cases,
having the concept of manually activated \texttt{tasks} and \texttt{stages}, case
worker activated events, and reacting to creation, update, and delete of case data via \texttt{entry
criteria} and \texttt{exit criteria}. Case planning itself can be modeled in advance, by providing
\texttt{tasks} and \texttt{stages} that allow planning with a
\texttt{planning table}. However, any worker in a role that allows planning can do planning at any
moment during the case instance execution. Planning is based on the concept of \texttt{discretionary
items} that are modeled to be used at the discretion of the case workers. Manually activated
\texttt{tasks} and \texttt{stages} are those for which the
\texttt{entry criterion} has been met, but they are only executed, if a case worker decides to do so.

CMMN formalizes the concept of a \texttt{case file} (or case folder) that contains all the
\texttt{case file items} (case data). In most instances, case data is represented by documents,
because workers commonly interact by using documents like spreadsheets, presentations, word processor documents, voice
recordings, videos, pictures, etc. The ability to add, modify, or remove data from a case file at any time during the
process is a key feature of a case management system.

Case workers always have access to the \texttt{case file} and all the case data. They can be authorized
to access the \texttt{case file} and its data even when they don't have work assigned. The interaction
of case workers with case data may trigger additional \texttt{tasks} or activities in the case.

{
\bibliographystyle{plain}
\bibliography{Tutorial}
}
\appendix
\clearpage
\section{CMMN version 1.0}
\subsection{Terminology}
\begin{description}
\item[Auto complete decorator] used to indicate that a \texttt{case plan} or \texttt{stage} will automatically complete when all the required \texttt{case plan items} have completed and there are no \texttt{case plan items} executing.
\item[Blocking human task] is a \texttt{task} that should be executed by a case worker and it will wait until the case worker completes it to be market as completed.
\item[Case file] is a container holding all the data and information generated during the execution of a case instance. Sometimes called a \texttt{case folder}.
\item[Case file item] is the data or information contained in a \texttt{case file}.
\item[Case instance] is the runtime representation of a \texttt{case plan} model.
\item[Case plan] is the set of \texttt{case plan items} that may be executed during the case instance execution. Note that a \texttt{case plan} model also describes the \texttt{discretionary items} that can be added to the plan at execution time.
\item[Case plan items] are \texttt{tasks}, \texttt{stages}, \texttt{event listeners}, and \texttt{milestones} that compose a \texttt{case plan}.
\item[Case plan model] is the model or diagram describing a case.
\item[Case planning] is that activity used by case workers to add \texttt{discretionary items} into the plan.
\item[Case task] is a \texttt{task} implemented by a case (modeled as a \texttt{case plan}).
\item[Case worker] is a person working in a case.
\item[Case] is synonymous with process.
\item[Connector] is a dotted line used in two situations. First, to indicate event propagation into an \texttt{entry criteria} or \texttt{exit criteria}'s \texttt{onPart}, in which case it is optional. Second, to indicate that a \texttt{discretionary item} belongs to a \texttt{planning table} of a \texttt{human task}.
\item[Discretionary item] is a \texttt{task} or \texttt{stage} that is not in the  plan, but it can be added to the  plan by a case worker doing planning.
\item[Entry criteria] describes a condition (\texttt{onPart} and \texttt{ifPart}) that is necessary for a \texttt{task}, \texttt{stage}, or \texttt{milestone} to execute. However, it may not be sufficient that the \texttt{entry criteria} is satisfied to execute a \texttt{task} or \texttt{stage} with a \texttt{manual activation} decorator.
\item[Event listener] represents an event that could occurs during the case instance.
\item[Exit criteria] describes a condition (\texttt{onPart} and \texttt{ifPart}) that will terminate the execution of a \texttt{task}, \texttt{stage}, or \texttt{case plan}.
\item[Human task] is a \texttt{task} executed by a case worker. There are two types of \texttt{human tasks}, namely \texttt{non-blocking human task} and \texttt{blocking human task}.
\item[IfPart] is an optional Boolean expression which is part of an \texttt{entry criteria} or \texttt{exit criteria}, and should evaluate to true for the \texttt{entry criteria} or \texttt{exit criteria} to be satisfied.
\item[Manual activation decorator] used to indicate that a \texttt{task} or \texttt{stage} may require a case worker to manually execute it.
\item[Milestone] represents an accomplishment during the case.
\item[Non-blocking human task] is a \texttt{task} that should be handed out to a case worker for execution, but it will not wait for the case worker to complete it to be marked as complete.
\item[OnPart] is an optional part of an \texttt{entry criteria} or \texttt{exit criteria}, and indicate which standard event can satisfy the entry criteria or exit criteria.
\item[Plan] is an execution concept. When a case instance start execution the plan contains all the non-discretionary \texttt{case plan items}. Case workers can add \texttt{discretionary items} to the plan.
\item[Plan fragment] is a set of \texttt{discretionary items} that can be added to the \texttt{case plan} in a single planning action.
\item[Planning table] is used to indicate when a \texttt{stage}, a \texttt{case plan}, or a \texttt{human task} contains \texttt{discretionary items}.
\item[Process task] is a \texttt{task} implemented by a process. The process may be modeled in any process or workflow modeling notation.
\item[Repetition decorator] used to indicate that a \texttt{task}, \texttt{stage}, or \texttt{milestone} with \texttt{entry criteria} may be executed multiple times (each time the \texttt{entry criteria} is satisfied).
\item[Required decorator] used to indicate that a \texttt{task}, \texttt{stage}, or \texttt{milestone} must be executed to consider the \texttt{case plan} or \texttt{stage} containing them completed.
\item[Role] describes what a case worker is allow to do. Each case worker is assigned to one or more roles in the case.
\item[Stage] is a composed activity that contains other \texttt{case plan items} in a case. It can be planned or discretionary.
\item[Standard event] is an event generated by either a \texttt{case file item} or a \texttt{case plan item}. CMMN has a list of standard events.
\item[Task] is an activity in a case. It can be planned or discretionary.
\item[Timer event listener] is an \texttt{event listener} that is configured to be satisfied by a timer.
\item[User event listener] is a \texttt{event listener} that expects to be satisfied by a human.
\end{description}
\clearpage
\subsection{Notational elements}
\begin{longtabu}{|ccp{7cm}|} \hline
\textbf{Planned} & \textbf{Discretionary} & \textbf{Description} \\\hline
\includegraphics[width=2cm]{case-plan-model.pdf}
& & case plan \\\hline

\includegraphics[width=2cm]{stage.pdf}
& \includegraphics[width=2cm]{discretionary-stage.pdf} 
& stage  \\\hline

\includegraphics[width=2cm]{task.pdf}
& \includegraphics[width=2cm]{discretionary-task.pdf} 
& task  \\\hline

\includegraphics[width=2cm]{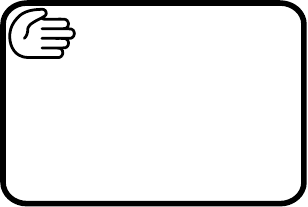} 
& \includegraphics[width=2cm]{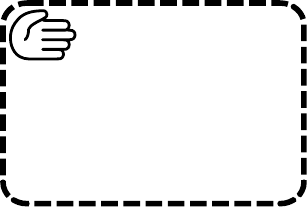}
& non-blocking human task  \\\hline

\includegraphics[width=2cm]{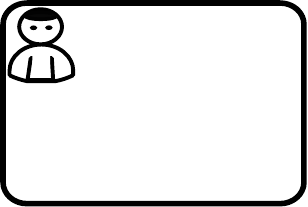} 
& \includegraphics[width=2cm]{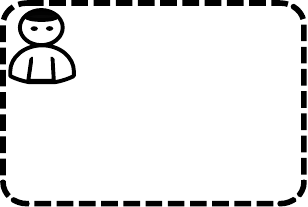}
& blocking human task   \\\hline

\includegraphics[width=2cm]{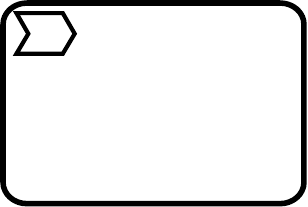} 
& \includegraphics[width=2cm]{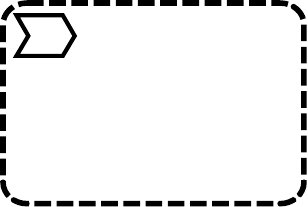}
& process task  \\\hline

\includegraphics[width=2cm]{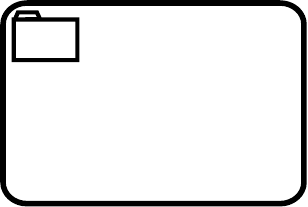}
& \includegraphics[width=2cm]{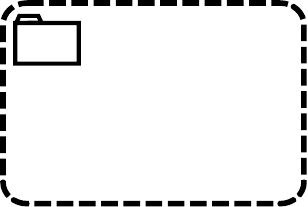}
& case task  \\\hline

~ & \includegraphics[width=2cm]{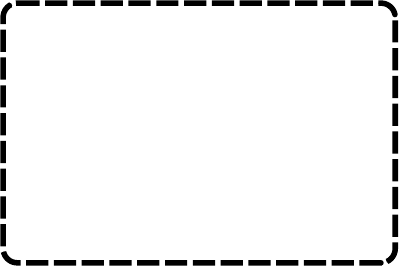} 
&  plan fragment  \\\hline

\includegraphics[width=1cm]{case-file-item.pdf}
& & case file item   \\\hline

\includegraphics[width=2cm]{milestone.pdf}
& & milestone   \\\hline

\includegraphics[width=0.7cm]{event-listener.pdf}
& & event listener   \\\hline

\includegraphics[width=0.7cm]{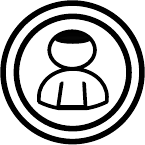}
& & user event listener   \\\hline

\includegraphics[width=0.7cm]{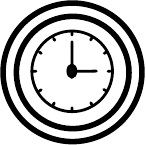}
& & timer event listener   \\\hline

$\ldots\ldots\ldots$ 
& & connector  \\\hline  
\end{longtabu}

\newgeometry{left=2cm}
\subsection{Applicability of Annotators}

\begin{center}
\small
\begin{longtabu}{|p{4cm}|c|c|c|c|c|c|c|c|} \hline
\textbf{Notational} & planning  & entry  & exit  & auto  & collapsed & manual  & repetition & required \\
~ &  table &  criterion &  criterion &  complete & expanded &  activation & ~ & ~ \\

& \includegraphics[width=0.4cm]{planing-table-nonvisualized.pdf}
& ${\lozenge}$ 
& ${\blacklozenge}$ 
& ${\blacksquare}$ 
& ${\boxplus}$ 
& ${\vartriangleright}$ 
& \texttt{\#} 
& $!$ \\

& \includegraphics[width=0.4cm]{planing-table-visualized.pdf}
& & & & ${\boxminus}$ 
& & &  \\\hline

\includegraphics[width=2cm]{case-plan-model.pdf}  & \checkmark & & \checkmark & \checkmark & & & & \\

case plan & & & & & & & & \\\hline

\includegraphics[width=2cm]{stage.pdf}  & \checkmark &  \checkmark &  \checkmark &  \checkmark &  \checkmark &  \checkmark &  \checkmark &  \checkmark \\

Stage  & & & & & & & & \\\hline

\includegraphics[width=2cm]{discretionary-stage.pdf}   & \checkmark &  \checkmark &  \checkmark &  \checkmark &  \checkmark &  \checkmark & \checkmark  &   \\

discretionary stage  & & & & & & & & \\\hline

\includegraphics[width=2cm]{task.pdf}  & &  \checkmark &  \checkmark & & &  \checkmark &  \checkmark &  \checkmark \\

task & & & & & & & & \\\hline

\includegraphics[width=2cm]{discretionary-task.pdf}   & &  \checkmark &  \checkmark & & &  \checkmark & \checkmark  &   \\

discretionary task & & & & & & & & \\\hline

\includegraphics[width=2cm]{non-blocking-human-task.pdf}   &  \checkmark &  \checkmark &   & & &  \checkmark &  \checkmark &  \checkmark \\

non-blocking human task & & & & & & & & \\\hline

\includegraphics[width=2cm]{non-blocking-human-discretionary-task.pdf} discretionary &  \checkmark &  \checkmark &   & & &  \checkmark & \checkmark  &   \\

non-blocking human task & & & & & & &  & \\\hline

\includegraphics[width=2cm]{blocking-human-task.pdf}   &  \checkmark &  \checkmark &  \checkmark & & &  \checkmark &  \checkmark &  \checkmark \\

blocking human task & & & & & & & & \\\hline

 \includegraphics[width=2cm]{blocking-human-discretionary-task.pdf} discretionary &  \checkmark &  \checkmark &  \checkmark & & &  \checkmark & \checkmark  &   \\

 blocking human task & & & & & & & & \\\hline

\includegraphics[width=2cm]{process-task.pdf}   & &  \checkmark &  \checkmark & & &  \checkmark &  \checkmark &  \checkmark \\

process task & & & & & & & & \\\hline

\includegraphics[width=2cm]{process-discretionary-task.pdf}  & &  \checkmark &  \checkmark & & &  \checkmark & \checkmark  &   \\

discretionary process task & & & & & & & & \\\hline

\includegraphics[width=2cm]{case-task.pdf}  & &  \checkmark &  \checkmark & & &  \checkmark &  \checkmark &  \checkmark \\

case task & & & & & & & & \\\hline

\includegraphics[width=2cm]{case-discretionary-task.pdf}  & &  \checkmark &  \checkmark & & &  \checkmark & \checkmark  &   \\

discretionary case task & & & & & & & & \\\hline

\includegraphics[width=2cm]{plan-fragment.pdf}   & & & & &  \checkmark & & & \\

plan fragment & & & & & & & & \\\hline

\includegraphics[width=1cm]{case-file-item.pdf}  & & & & & & & & \\

case file item & & & & & & & & \\\hline

\includegraphics[width=1.8cm]{milestone.pdf}  & & \checkmark & & & & & \checkmark  &  \checkmark \\

milestone & & & & & & & & \\\hline

\includegraphics[width=0.7cm]{event-listener.pdf}  & & & & & & & & \\

event & & & & & & & & \\\hline

\includegraphics[width=0.7cm]{event-listener-user.pdf}  & & & & & & & & \\

user event & & & & & & & & \\\hline

\includegraphics[width=0.7cm]{event-listener-timer.pdf}  & & & & & & & & \\

timer event & & & & & & & & \\\hline

\end{longtabu}
\end{center}

\clearpage
\subsection{Standard CMMN Events}
\begin{longtabu}{|c|l|p{12cm}|} \hline
\textbf{Notation} & \textbf{Events}& \textbf{Description} \\\hline
\multicolumn{3}{|c|}{\textbf{Plan Items} }\\\hline
\multirow{7}{*}{\includegraphics[width=2cm]{case-plan-model.pdf}} & \multicolumn{2}{p{12cm}|}{\textbf{Case events}} \\\cline{2-3}
~ & \textbf{create} & The case was created. \\
~ & \textbf{suspend} &   The case was Suspended. Case workers can suspend a case. \\
~ & \textbf{reactivate} &   The case was reactivated. A case that is suspended, completed, terminated or in fault can be reactivated. A case worker can reactivate a case. \\
~ & \textbf{complete} &  The case completed normally, all required tasks and stages completed.\\
~ & \textbf{terminate} &  The case reached an exit criteria or a case worker terminated the case.  \\
~ & \textbf{fault} &  The case entered in a failed state.\\
~ & \textbf{close} &  The case was closed by a case worker.  A case that is suspended, completed, terminated or in fault can be closed, and it cannot be reactivated again.\\\hline

~ & \multicolumn{2}{p{12cm}|}{\textbf{Task and Stage events}} \\\cline{2-3}
\multirow{2}{*}{\includegraphics[width=2cm]{task.pdf}} 
 & \textbf{create} & The task or stage is created. This happens when the stage or case containing them start executing. Or when a case worker adds a discretionary task or discretionary stage to the plan.\\
~ & \textbf{start} &  When an entry criteria for a task or stage without the `manual activation' flag is satisfied the task or stage start executing.\\
\multirow{2}{*}{\includegraphics[width=2cm]{discretionary-task.pdf}}
 & \textbf{enable} &  When an entry criteria for a task or stage with a `manual activation' flag is satisfied the task or stage becomes enabled.\\
~ & \textbf{manualStart} &  An enabled task or stage must be manually activated by a case worker to start executing.\\

\multirow{2}{*}{\includegraphics[width=2cm]{stage.pdf}}   

  & \textbf{disable} &  A stage or task that is enabled can be disabled by a case worker.\\
~ & \textbf{reenable} &  A stage or task that was disabled can be enabled again by a case worker.\\

 & \textbf{suspend} &   A case worker can suspend a task or stage. \\
 & \textbf{resume} &   A case worker can resume a task or stage that was suspended.  \\

\multirow{2}{*}{\includegraphics[width=2cm]{discretionary-stage.pdf}}  & \textbf{parentSuspend} &  A stage or task is suspend by its parent stage or case when that parent stage or case is suspended by a case worker.\\
~ & \textbf{parentResume} &  A stage or task is resumed by its parent stage or case when that parent stage or case is resumed by a case worker.\\

~ & \textbf{fault} &  The stage or task transitions has reached a failure condition.\\
~ & \textbf{reactivate} &   A task or stage that is in a failure condition can be reactivated by a case worker.\\

~ & \textbf{exit} &  The Stage has reach an exit criteria.\\
~ & \textbf{complete} &  The task or stage has completed normally. \\
~ & \textbf{terminate} & The task or stage is terminated by a case worker or when the stage or case containing it terminates.  \\\hline

~ & \multicolumn{2}{p{12cm}|}{\textbf{Events and Milestone events}} \\\cline{2-3}
\multirow{2}{*}{\includegraphics[width=0.7cm]{event-listener.pdf}}
~ & \textbf{create} & The event or milestone is created.  This happens when the stage or case containing them start executing.\\
\multirow{2}{*}{\includegraphics[width=1.8cm]{milestone.pdf}} 

 & \textbf{suspend} &   The event or milestone is suspended. A case worker can suspend an event or milestone, but they are also suspended when the case or the stage containing them is suspended.\\
~ &  \textbf{resume} &   An event or milestone that is suspended can resume. A case worker can resume an event or milestone, but they are also resumed when the case or the stage containing them is resumed\\
~ & \textbf{occur} &  The event occurs or the milestone is reached.\\
~ & \textbf{terminate} &  The event or milestone is terminated by a case worker or when the stage or case terminates.\\\hline

\multicolumn{3}{|c|}{\textbf{Data (Case file items) events}}\\\hline
\multirow{5}{*}{\includegraphics[width=0.9cm]{case-file-item.pdf}} &
\multicolumn{2}{p{15cm}|}{All type of data (for example, a row in a database, a document, a picture, a video, etc.) have the following events}\\\cline{2-3}
~ & \textbf{create} & The item is created. \\
~ & \textbf{replace} & The content of the item has been replaced.\\
~ & \textbf{update} & The item has been updated. \\
~ & \textbf{delete} & The item has been deleted.\\
~ & \textbf{addReference} & A new reference to the item has been added. \\
~ & \textbf{removeReference} & A reference to the item has been removed. \\\cline{2-3}
~ & \multicolumn{2}{p{15cm}|}{In addition to basic data events, containers (for example, a directory, a folder, a set, a stack, a list, etc.) have two more events} \\\cline{2-3}
~ & \textbf{addChild} & A new child  has been added to an existing case file item. \\
~ & \textbf{removeChild} & A child has been removed from a case file item. \\\hline
\end{longtabu}

\end{document}